\documentclass[
pre, 
twocolumn, superscriptaddress, a4paper, showpacs,
]{revtex4}
\usepackage{graphicx}
\usepackage{amsmath}

\makeglossary

\begin{document}

\title{Boundary-reaction-diffusion model for oscillatory zoning\\ in binary crystals grown from solution}
\date{\today}

\author{Felix Kalischewski}
 \email{Kalischewski@uni-muenster.de}
 \affiliation{Westf\"alische Wilhelms
Universit\"at M\"unster, Institut f\"ur physikalische Chemie, Corrensstr. 30,
48149 M\"unster, Germany}
 \affiliation{Center of Nonlinear Science CeNoS,
Westf\"alische Wilhelms Universit\"at M\"unster, 48149 M\"unster,
Germany}
\author{Ihor Lubashevsky}
 \email{ialub@fpl.gpi.ru}
 \affiliation{Westf\"alische Wilhelms Universit\"at
M\"unster, Institut f\"ur physikalische Chemie, Corrensstr. 30, 48149
M\"unster, Germany}
 \affiliation{A.M. Prokhorov General Physics Institute, Russian Academy of
Sciences, Vavilov Str. 38, Moscow, 119991 Russia}
\author{Andreas Heuer}
 \email{andheuer@uni-muenster.de}
 \affiliation{Westf\"alische Wilhelms
Universit\"at M\"unster, Institut f\"ur physikalische Chemie, Corrensstr. 30,
48149 M\"unster, Germany}
 \affiliation{Center of Nonlinear Science CeNoS,
Westf\"alische Wilhelms Universit\"at M\"unster, 48149 M\"unster, Germany}

\begin{abstract}
Oscillatory Zoning (OZ) is a phenomenon exhibited by  many geologically formed
crystals. It is characterized by quasi periodic oscillations in the composition
of a solid solution, caused by self-organization. We present a new model for
OZ. The growth mechanism applied includes species diffusion through the
solution bulk, particle adsorption, surface diffusion and subsequently
desorption or incorporation into the crystal. This mechanism, in particular,
can provide the synchronization effects necessary to reproduce the layered
structure of experimentally obtained crystals, lacking in other models. We
conduct a linear stability analysis combined with numerical simulations. Our
results reproduce the experimental findings with respect to the patterns formed
and a critical supersaturation necessary for OZ to occur.
\end{abstract}


\maketitle

\section{Introduction}
Oscillatory zoning (OZ) is a phenomenon describing repetitive composition
variations of binary solid solutions along their core-to-rim profile.
Traditionally, it was believed to be of rare occurrence and its existence was
ascribed to variations of external parameters controlling the crystal growth,
like temperature or concentration fluctuations. However, the development of
more sophisticated observation techniques facilitated the detection of this
phenomenon in all major classes of minerals and a wide range of geological
environments \cite{ShoreFowler}. In addition to naturally obtained samples OZ
was experimentally reproduced in the absence of external fluctuations. Reeder
\textit{et al.} \cite{OZMnExp} were able to grow calcite crystals exhibiting OZ
of the Mg dopant and Putnis \textit{et al.}
\cite{PrietoZoningExp,PutnisNature,PinaComposition} obtained end-member zoning
in (Ba,Sr)SO$_4$ crystals.

\begin{figure}[h]
\center
\includegraphics[width=0.75\linewidth]{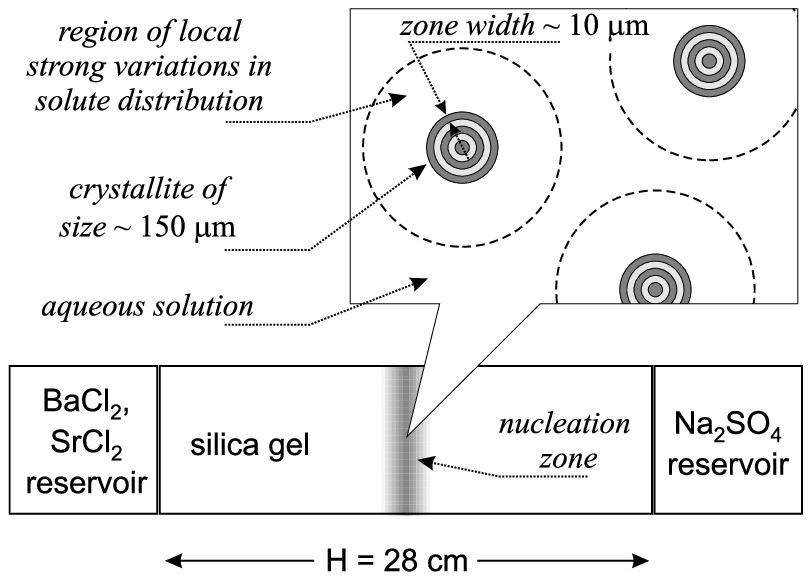}
\caption{\label{ExpPutnis} Experimental setup in which oscillatory zoned
crystals of (Ba,Sr)SO$_4$ were synthesized in
Refs.~\cite{PrietoZoningExp,PutnisNature,PinaComposition}. The reactants
counterdiffuse in the column and (Ba,Sr)SO$_4$ crystals nucleate. The upper
window sketches the structure of the nucleation zone and the length scales
involved.}
\end{figure}

The experimental setup used by Putnis \textit{et al.}
\cite{PrietoZoningExp,PutnisNature,PinaComposition} is sketched in
Fig.~\ref{ExpPutnis}. It consists of two reservoirs, one filled with an aqueous
solution of BaCl$_2$/SrCl$_2$ and the other with Na$_2$SO$_4$. The two
reservoirs are connected by a column filled with a silica-gel to inhibit
convective transport. With the beginning of the experiment the reactants from
the reservoirs start to diffuse toward each other through the column. As the
diffusion fields of Ba$^{2+}$ and Sr$^{2+}$ from one reservoir and
SO$_4{}^{2-}$ from the other reservoir exceed the nucleation threshold product
in the vicinity of the column center, nuclei form. The solution is then
strongly supersaturated with respect to the freshly generated crystal seeds and
the growth commences in a layer of few  millimeters in width
\cite{PutnisPrivComm}. After approximately one month the experiment was
terminated. The obtained crystals exhibited OZ although no external
fluctuations were imposed on the system. Thus it has been clearly shown that OZ
can be also attributed to intrinsic mechanisms resulting in spontaneous
structure formation \cite{Ortoleva}. The wide range of different crystals
concerned suggests a certain universality of the underlying mechanism.

The general principle causing OZ is the autocatalytic or inhibiting interaction
of the substrate with the end member concentration in melt or solution
\cite{OrtolevaFeedback}. If, for example, the crystal is rich in component A,
this will lead to increased growth of this component in an autocatalytic way.
Its supply will eventually be limited by diffusion. During this phase the
disfavored component B will accumulate in the solution, leading to a slight
increase of B deposition. However, any small increase in B will show
autocatalytic effects, whereas the growth of A slowly decreases. The
combination of a relatively high B concentration in combination with the
autocatalytically increasing growth rate will then lead to a phase of B
dominated growth. With this, A and B have switched roles and one half cycle is
completed. If the interaction of those two processes is interrupted, for
example by stirring no OZ will be observed \cite{OZMnExp}.

The specific interaction of autocatalytic growth and component accumulation is
subject to the scenario employed and gives rise to different nonlinear schemes.
The first quantitative model derived by Haase \textit{et al.} \cite{Haase}
describes self-organized oscillatory zoning from the melt applying moving
boundaries and a generically autocatalytic growth term. In a subsequent series
of publications Wang and Merino introduced the boundary layer approximation for
the treatment of zoned crystals grown hydrothermally
\cite{WangMerinoHydrothermal}, from solution \cite{WangMerinoSolution}, and
from melt \cite{WangMerinoMelt}. Later, non generic growth terms derived from
the physics of growth processes from the melt were introduced by L'Heureux
\cite{HeureuxMeltHeatDiff} using constitutional undercooling and by Wang and Wu
\cite{WangWu} employing the excess enthalpy of crystallization.

The most sophisticated models currently available for end member OZ from
solution have been developed by L'Heureux \textit{et al.}
\cite{PhanMathMod,KatsevNoise,HeureuxCellAuto}. These models apply the boundary
layer approximation, as well, and in addition to the otherwise deterministic
nature the influence of noise on the onset of OZ is studied. The non-linear
growth term applied is phenomenologically obtained from the local probability
to find a matching kink site as proposed by Markov \cite{MarkovBuch}. However,
the local nature of this mechanism does not provide the synchronization effects
necessary to describe homogeneous growth fronts resulting in the ring like
composition oscillations found in the experiment \cite{HeureuxCellAuto}.

In this paper we present a boundary reaction diffusion model for OZ in binary
solid solutions grown from aqueous solution. We abandon the boundary layer
approximation and explicitly treat the diffusion above the crystal without
further approximations. Furthermore, the growth rate, being the central
ingredient of every model, is derived directly from considerations of the
physical growth mechanisms. We apply the concept of layer-by-layer growth under
continuous generation of new steps which is, e.g., relevant for growth by screw
dislocations or 2D nucleation. The growth mechanism results as an interplay of
different processes including bulk diffusion, adsorption, surface diffusion,
and eventually desorption or incorporation into the crystal. The non-linearity
necessary to generate OZ is obtained by the composition-dependence of the mean
life time of adatoms in the adsorbed layer or, equivalently, the interaction of
adatoms with the crystal surface.

In Sect. II the different aspects of the model are introduced. The resulting
model equations are summarized in Sect. III. Then, Sect. IV analyses this model
close to the stationary point. Sect. V presents its numerical analysis,
including all nonlinear effects. Finally, the obtained results are discussed in
Sect. VI and concluded in Sect. VII.

\section{Physical background}
The model under consideration describes the diffusion processes in the bulk
solution, the growth process following from the coupling between crystal and
solution, and the evolution of the crystal composition.

\subsection{General}

\begin{figure}[h]
\center
\includegraphics[width=0.75\linewidth]{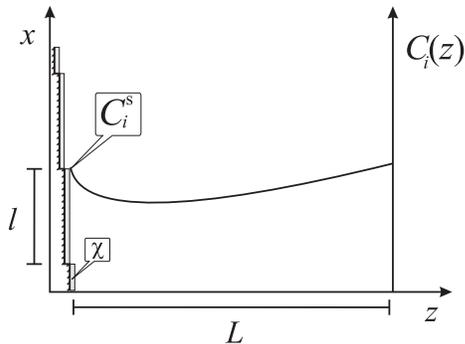}
\caption{\label{ModelScheme}Scheme of the 2D-model. The concentration is
averaged over a sufficiently large $x$-region, giving rise to a 1D-model with a
concentration $C(z)$.}
\end{figure}

Based on the slow crystal growth observed in the experiments
\cite{PrietoZoningExp,PutnisNature,PinaComposition}, we apply screw
dislocations as the step generating mechanism and describe the
subsequent growth process via step advance. Screw dislocation driven
growth can cross over to two dimensional nucleation, as shown by
Pina \textit{et al.}~\cite{PinaComposition}. However, this will not
affect the validity of the model because the specific process of
step generation is not of immediate importance. We assume that after
nucleation the crystal surface will reach a steady-state when the
density of the step generating islands or spirals does not change {
any more, because of coalesence of the terraces according to
\cite{BCF}}. A refined model would be necessary to take into account
any effects related to anisotropic growth as found by Pina
\textit{et al.} \cite{PinaNature} or to account for the curvature of
small spirals. Since we are interested in the basic mechanisms we
refrain from such detailed modelling and just consider infinite step
trains which on average are a distance $l$ apart; see
Fig.~\ref{ModelScheme}. Typically, $l$ is in the nanometer-regime.
The coordinates $z$, characterizing the distance from the crystal
surface and $x$, orthogonal to the steps, are indicated. This
describes a one-dimensional crystal surface which starts at $z=0$,
i.e. the total system is a 2D-system.

In general, quantities like the solute concentration $C(x,z)$ depend on
$x$ and $z$. Conceptually, the $x$-dependence can be separated into two
different contributions. First, there exists a periodic contribution with
quasiperiod $l$. It expresses the fact that the concentration close to the
steps will be smaller. However, in the limit, considered below, only the
concentration, averaged over the length scale of $l$, will enter. Therefore
this $x$-dependence of $C(x,z)$ is not relevant. Second, there can be
variations on length scales much larger than $l$. Experimentally, it is
observed that the growth behavior does not change along the surface of one
crystallite of size 150~$\mu$ at a given time. Furthermore, a straightforward
extension of the stability analysis, presented below, shows that the maximum
instability for fluctuations along one crystal surface are for zero wave
vector. Thus it is realistic to hope that the leading mechanism of OZ can be
derived from study of the $z$-dependence alone. In the present work, we will
therefore neglect a possible long-range $x$-dependence and restrict ourselves
to the 1D model.

In the present paper we deal with a two species model ($i=1,2$) for
the crystal growth from solution (Ba and Sr, respectively), thereby
neglecting possible variations of the SO$_4{}^{2-}$ concentration.
This can be justified in two ways. First, using the OZ model by
L'Heureux \cite{PhanMathMod} we have verified that { a system with
artificially fixed} SO$_4{}^{2-}$ concentration exhibits basically
the same dynamics \cite{Diplomarbeit}. Second, it describes also the
solid state formation for the three component system SO$_4{}^{2-}$,
Ba, and Sr, studied by Putnis \textit{et al.} when the concentration
of SO$_4{}^{2-}$ is sufficiently high.  The mole fraction of
component 1 on the crystal surface is denoted by $\chi$ ($0 \le \chi
\le 1$). { Throughout this work we assume the same molar volume for
both species in the solid phase}.

\subsection{Crystal-Solution Interface}
The central element of this model is the coupling term between the crystal
growth rate and the surface species concentration. We have used the classical
approach by Gilmer \textit{et al.} \cite{GilmerSurfaceAndVolumeDiffusion} which
itself builds on the BCF-theory \cite{BCF}: solute particles adsorb on the
crystal surface and diffuse along it. If before desorbing they come into
contact with a step on the crystal surface, they are incorporated. If they do
not meet a step in time, they desorb and become part of the solution again.

On their path along the crystal surface the adatoms experience many different
local environments depending on the crystal composition. The mean energy
difference $E_i(\chi)$ of the adsorption reflects the interaction with the
crystal surface and the solvation process. Within the mean-field description it
can be approximated by a linear combination of the different adsorption
energies:
\begin{align}
\label{MeanEnergy}
E_{1}(\chi)&=\chi\,E_{11} + (1-\chi)\,E_{12} - E_1^\mathrm{sol}\,,\nonumber\\
E_{2}(\chi)&=\chi\,E_{12} + (1-\chi)\,E_{22} - E_2^\mathrm{sol}\,.
\end{align}
Here, $E_{ij}=E_{ji}$ is the adsorption energy for component $i$ on a surface
formed by component $j$ and $E_i^\mathrm{sol}$ represents the solvation energy
of species $i$.

In the subsequent mathematical treatment it is useful to rewrite
expressions \eqref{MeanEnergy} in the symmetrized form
\begin{align}
E_1(\chi) &= 2\left(-\chi\phi + (1-\chi)\theta + \eta \right)k_BT + \Delta E^\mathrm{ad}_\mathrm{m}\,, \nonumber\\
E_2(\chi) &= 2\left(\chi\theta + (1-\chi)\phi - \eta \right)k_BT + \Delta
E^\mathrm{ad}_\mathrm{m}\,, \label{SymmetricEnergy}
\end{align}
where the dimensionless potentials
\begin{align}\label{ReducedVariables}
\phi   &= \left(E_{22}-E_{11}\right)/4k_BT\,, \nonumber\\
\theta &= \left(2E_{12}-\left(E_{11}+E_{22}\right)\right)/4k_BT\,, \nonumber\\
\eta   &= \left(E_{2}^\mathrm{sol}-E_{1}^\mathrm{sol}\right)/4k_BT\,.
\end{align}
and the mean { homogeneous adsorption energy}
\begin{equation}
\Delta E^\mathrm{ad}_\mathrm{m} = (1/2) (E_{11} + E_{22}) - (1/2)
(E_1^\mathrm{sol} + E_2^\mathrm{sol})
\end{equation}
have been introduced. Here, $k_B$ is the Boltzmann constant and
$T$ the temperature.  The potential $\phi$ represents the
asymmetry between homogeneous adsorption energies, whereas
$\theta$ is a measure for the preference of homogeneous over
heterogeneous adsorption. We consider the case $0 \le \phi \le
\theta$. The potential $\phi$ can be assumed to be nonnegative due
to symmetry reasons; see Eq.~\eqref{ReducedVariables}.   The limit
$\phi=\theta$ implies $E_{12} = E_{22}$, so that the crystal
growth properties of species 2 are independent of the composition
of the crystal surface. The case $\phi
> \theta$, corresponding to a different type of  crystal growth instability,
is beyond the scope of the present paper.  The last parameter $\eta$
reflects the solution energy difference of the two types of
particles.

For the crystal growth two time scales are of primary importance. First, the
inverse of the adsorption time $\tau_a$ denotes the rate with which a particle
in the solution layer above the crystal surface adsorbs. Thus, $a C_i^s/\tau_a$
is the particle flux on the crystal surface where $a$ is the typical distance
between atoms and $C_i^s =C_i(z=0)$ the mean concentration of component $i$ in
the solution just above the crystal surface. We assume that $\tau_a$ is the
same for both species.  Second, $\tau_{d,i}$ is the mean residence time of
adatoms on the surface. Detailed balance between both time scales requires
\begin{equation}
\tau_{d,i}(\chi) = \tau_a \exp[-E_{i}(\chi)/k_BT]\,.
\end{equation}
Thus the composition-dependence of $\tau_{d,i}$ is due to the
composition-dependence of the adsorption potentials $E_i(\chi)$. From
$\tau_{d,i}$ the mean diffusion length $l^\mathrm{s}_{i}$ can be obtained using
the Einstein relation
\begin{align}\label{SurfaceDiffusionLength}
l^\mathrm{s}_{i}(\chi)  &=\sqrt{D_s\cdot\tau_{d,i}(\chi)}\,,\nonumber\\
                        &=\sqrt{D_s \cdot \tau_a} \exp\left[-\frac{1}{2}\;
                        E_i(\chi)/k_BT\right].
\end{align}
The adatom diffusion coefficient $D_s$ characterizes the elementary
atomic movements of the adatoms on the crystal surface. We also
assume that $D_s$ is the same for both the species.

By substitution of equation~\eqref{SymmetricEnergy} one explicitly obtains for
the mean diffusion length
\begin{align}
l^\mathrm{s}_{1}(\chi)  &= l_D\;f_\eta\;\exp\left[-\left( -\chi\phi + (1-\chi)\theta\right)\right]\,,\nonumber\\
l^\mathrm{s}_{2}(\chi)  &= l_D\;f_\eta^{-1}\;\exp\left[-\left(\chi\theta +
(1-\chi)\phi\right)\right]\\
  \intertext{with}
l_D &:= \sqrt{D_s \tau_a} \; \exp(-\Delta E^\mathrm{ad}_\mathrm{m}/2k_BT)\\
 \intertext{and}
 \label{fetadef}
 f_\eta &:= \exp(-\eta).
\end{align}

Now the  partial growth rate $r_i$ of species $i$ can be expressed
as a combination of the adsorption flux and a success factor $q_i$,
describing the fraction of adatoms that will actually contribute to
crystal growth, whereas the others desorb:
\begin{equation}
r_i(\chi,C^\mathrm{s}_i)=\underbrace{\frac{a}{\tau_a}\cdot
C^\mathrm{s}_i}_\text{flux to surface}{}\cdot q_i(\chi)\,.
\end{equation}
Following Ref.~\cite{PhanMathMod} we have neglected a term, representing the
equilibrium concentration which is irrelevant under significant supersaturation
as present in the experiment. The adsorption of a particle is hindered by the
breakup of the solution shell and consequently takes longer than a normal
diffusion step in the solution bulk. It is reasonable to assume that the
adsorption time scale $\tau_a$ is much larger than the typical time scale
$a^2/D$ of free diffusion in the solution, i.e. $\sqrt{D\,\tau_a}\gg a$. Since
$l$ is also a microscopic length scale one may even expect that
$\sqrt{D\,\tau_a}\gg l$. Then the location of successive adsorption processes
of the same particle are uncorrelated with respect to the position of the steps
and this simple probabilistic approach becomes justified.

The growth is supported by surface diffusion from both the sides of a step. In
the limit $l^\mathrm{s}_{i} \ll l$ the atoms adsorbing within the distance
$l^\mathrm{s}_{i}$ of a step will typically contribute to crystal growth,
whereas all the remaining adatoms will desorb after some time. Therefore, the
probability to meet a step is of the order of $l^\mathrm{s}_{i}/l$ and can more
precisely be expressed as \cite{GilmerSurfaceAndVolumeDiffusion}
\begin{equation} \label{GrowthRateLimit}
q_i(\chi) = \frac{2\,l^\mathrm{s}_{i}(\chi)}{l} \ .
\end{equation}
For arbitrary ratio $l^\mathrm{s}_{i} / l$  one obtains an additional factor
$\tanh (l/2\,l^\mathrm{s}_{i})$ which in the limit $l^\mathrm{s}_{i} \ll l$
turns out to be unity \cite{GilmerSurfaceAndVolumeDiffusion}.

For later convenience we gather the system constants into a
characteristic time
\begin{equation}
\label{tausimpledef} \tau=\frac{l\,\tau_a}{2\,l_D}
\end{equation}
and defining
\begin{equation}
\label{tauidef}
 \tau_i(\chi) = \tau \frac{l_D}{l^\mathrm{s}_{i}(\chi)}
\end{equation}
the individual growth rates can be written as
\begin{equation}
\label{rate} r_i(\chi,C_i^s) = \frac{a C^\mathrm{s}_i}{\tau_i(\chi)}\,.
\end{equation}
Then the resulting continuity relation at the crystal surface reads
\begin{equation}
\label{bound0}
 D_i\left. \frac{\partial C_i(z,t)}{\partial z}
\right|_{z=0} = r_i(\chi,C_i^s)\,.
\end{equation}
This expresses the particle flux, on the left side, from the solution to the
crystal and, on the right side, in the continuous solution close to the
surface. An analogous boundary condition has been used in the classical work by
Gilmer \textit{et al.} \cite{GilmerSurfaceAndVolumeDiffusion}.

\subsection{Bulk Solution}
{The description  of the bulk solution follows the following
picture: At the beginning of the experiment the components begin to
diffuse from the reservoirs into the gel column. As their diffusion
profiles begin to overlap close to the middle of the column and the
nucleation barrier is overcome, crystallites form. Due to the the
narrow nucleation zone \cite{PutnisPrivComm} these nuclei must act
as an effective sink with respect to the current of Ba$^{2+}$,
Sr$^{2+}$, and SO${_4}^{2-}$ from their reservoirs.

Based on these considerations, this model is set up as a
source-sink-system with a gradient in between. In this aspect it
differs distinctly from the models proposed by L'Heureux \textit{et
al.}\cite{PhanMathMod,KatsevNoise,HeureuxCellAuto}, where the
crystal is considered to be growing through a homogeneously
supersaturated medium. Consequentially an analogous boundary layer
approximation cannot be applied to the present system.}

Experimentally, the following scales are observed \cite{PutnisNature}: (i)
Growth velocity  $V \approx 10^{-8}$~cm/s as estimated from the crystal
thickness and the total growth time. (ii) Thickness $H_{oz} \approx 10^{-5}$~cm
of individual layers. (iii) Time $T_{\rm oz} \approx H_{\rm oz}/V \approx 10^5$~s during which one pattern layer is formed. (iv) Bulk diffusion constant $D
\approx 10^{-5}$~cm$^2$/s.

>From these observables two important length scales can be estimated.
(i) The length scale $L_{\rm oz}$ characterizes the spatial
variations of the species distribution caused by oscillatory zoning.
It is given by $L_{\rm oz} \approx (DT_{\rm oz})^{0.5} \approx 1$~cm
which is much larger than the crystal size.  This is consistent with
the fact that the growth behavior does not change along one crystal
surface, because during times of small change in $\chi$ the
information about the local surface concentration can spread over
the whole region. Besides, it rationalizes the mean field approach
discussed below. (ii) Furthermore, one may wonder to which degree
the motion of the solid phase boundary caused by crystal growth can
affect the diffusion fields. For small length scales the
concentration field is determined by diffusion ($\approx
(Dt)^{0.5}$) whereas for large scales it is determined by the
(nearly) constant growth of the phase boundary ($\approx vt$). The
crossover length scale $L_{\rm v}$ for which both processes are
equally relevant can be thus estimated as $L_{\rm v} \approx D/V$
which for the present situation is close to $10^3$~cm. This scale
exceeds the system size by orders of magnitude.

Because of this estimate the effect of the growth induced motion of the crystallites boundaries on the solute diffusion is
ignorable and the crystal surface can be regarded as fixed with respect to the mathematical model.

Finally, we choose the external boundary condition such that the
reservoir is characterized by a constant influx $G_i$ of solute
into the system at $z = L$ where $L$ is an arbitrary large length
scale.

\subsection{Surface composition evolution}
To complete the model, a governing equation for the evolution of the crystal
surface composition is necessary. Assuming a homogeneous distribution of the
components throughout the surface, the composition change with the next time
increment $dt$ can be expressed as a function of the current composition $\chi$
and the relation of the growth rates $r_i$:
\begin{equation}\label{chidot1}
\frac{d\chi}{dt}=a^2\,\Bigl[\left(1-\chi\right)\cdot r_1(\chi,C_1^s)-\chi\cdot
r_2(\chi,C_2^s)\Bigr]\,.
\end{equation}
This relation reflects mass conservation and has already been  used in previous
work on OZ \cite{PhanMathMod}.

\section{The Boundary-Reaction-Diffusion Model}
\label{modelall}
 The 1D formulation of the model discussed above can be described as
 follows. Diffusion of the components $i=1,2$ through the
solution is considered within the region $z \in [0,L]$ and is
described by the equation
\begin{equation}\label{Diffusion}
\frac{\partial C_i(z,t)}{\partial t} = D_i\frac{\partial^2 C_i(z,t)}{\partial
z^2}\,,
\end{equation}
where $L$ should be chosen large enough to satisfy $\partial C/\partial t = 0$.
At the external boundary $z=L$ the influx of both the components is fixed
\begin{equation}\label{Influx}
G_i = D_i\left. \frac{\partial C_i(z,t)}{\partial z} \right|_{z=L}.
\end{equation}
At the crystal surface ($z=0$) the diffusion flux and the rate of the crystal
growth are related by (using expressions \eqref{rate} and \eqref{bound0})
\begin{equation}
\label{boundaryfin} D_i\left. \frac{\partial C_i(z,t)}{\partial z}
\right|_{z=0} = \frac{aC^\mathrm{s}_i}{\tau_i(\chi)}\,.
\end{equation}
The time scales $\tau_i$, as defined in Eq.\eqref{tauidef}, are given by
\begin{align}
\tau_1(\chi) &= \tau f_\eta^{-1} \exp\left[-\chi (\phi+\theta ) +
\theta\right]\,,
\nonumber\\
\tau_2(\chi) &= \tau\,f_\eta \exp\left[\chi(\theta-\phi)+\phi
\right].\label{taudef}
\end{align}
The constants $\tau$ and $f_\eta$ have been defined in
Eqs.~\eqref{tausimpledef} and \eqref{fetadef}, respectively. Finally, for the
solid composition evolution one obtains (compare Eq.~\eqref{chidot1})
\begin{equation}\label{chidot}
\frac{d\chi}{dt}=a^2\,\Bigl[\left(1-\chi\right)\,\frac{a
C^s_1}{\tau_1(\chi)}-\chi\,\frac{a C^s_2}{\tau_2(\chi)}\Bigr]\,.
\end{equation}

In order to complete the description of OZ it is necessary to calculate the
resulting structure of the crystal, i.e. $\chi_{\rm crystal}(z_{ \rm
crystal})$. For this purpose we introduce $\zeta(t)$ as the location of the
crystal surface at time $t$ in a coordinate system which does not move with the
crystal surface. Defining $t_z$ as the time for which $\zeta(t) = z_{\rm
crystal}$ one has
\begin{equation}
\chi_{\rm crystal}(z_{\rm crystal}) = \chi(t_z)\,.
\end{equation}
The function $\zeta(t)$ can be easily obtained from
\begin{equation}
d\zeta/dt  = a^3\left[\frac{a C^s_1}{\tau_1(\chi)}+\frac{a
C^s_2}{\tau_2(\chi)}\right],
\end{equation}
where the left side can be interpreted as the time-dependent growth velocity
proportional to the cumulative species flux at the surface.

\section{Stability Analysis}

The following two sections analyze the behavior of the  crystal
growth within the boundary-reaction diffusion model specified in the
previous one. Particularly we study the crystal growth properties
close to the stationary point by means of linear stability analysis.

\subsection{Stationary solution}

Both, Eqs.~\eqref{Diffusion} and  \eqref{chidot} describe the time-dependence
of the underlying system. For each equation and for fixed $\chi$ the
disappearance of the time-derivative yields some ratio $C_1^s/C_2^s$,
respectively.

\begin{figure}
\begin{center}
\includegraphics[width = 0.8\columnwidth]{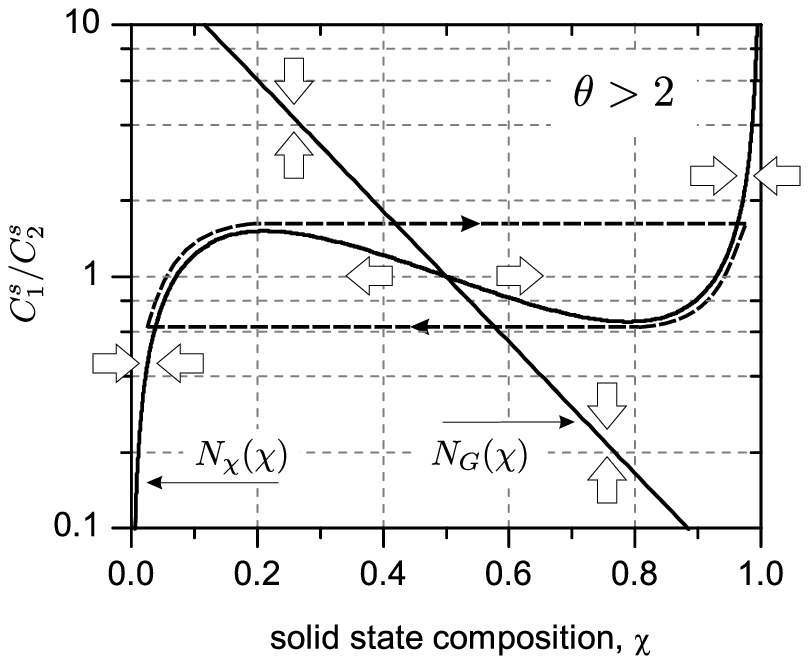}
\end{center}
\caption{Stationary functions $N_\chi(\chi)$ and $N_G(\chi)$ vs.
the solid state composition $\chi$. The chosen parameters are
$\theta = 3$, $f^2_\eta e^\phi = 1$, and $G_1 = G_2$. For this
choice the stationary value $\chi^{st}$ is 0.5. The dashed curve
illustrates the typical construction of the limit cycle for the
developed instability.\label{F:LSA:1}}
\end{figure}

Evidently, $(d/dt)\chi(t) = 0$ implies { (using $f_\eta^2 e^\phi=1$
for simplicity)}
\begin{equation}
\label{stat0} \frac{C_1^s}{C_2^s} = \frac{\chi}{1-\chi}
\frac{\tau_1(\chi)}{\tau_2(\chi)} = \frac{\chi}{1-\chi}
\exp(\theta(1-2\chi))\equiv N_\chi(\chi)\,.
\end{equation}
This nullcline $N_\chi(\chi)$ is shown in Fig.~\ref{F:LSA:1}. It possesses a
decreasing branch located in the region $\chi\in\big( \chi_{-},\chi_{+}\big)$
\begin{equation}
    \chi_\pm: = \frac12\left(1\pm\sqrt{\frac{\theta-2}{\theta}}\right),
\end{equation}
when the parameter $\theta$ exceeds the critical value $\theta_c =
2$. For $\theta < 2$  the function $N_\chi(\chi)$ is monotonously
increasing. In the stationary case the presence of a decreasing
branch implies that different values of $\chi$ are associated  to
the same $C_1^s/C_2^s.$

A stationary solution of the diffusion equation Eq.~\eqref{Diffusion},
fulfilling the boundary conditions Eqs.~\eqref{boundaryfin}, reads
\begin{equation}
    \label{EVQ:6}
      C_i(z) = \frac{G_i \tau_i(\chi)}{a} + \frac{G_i}{D_i} z \,,
\end{equation}
giving the surface concentration
\begin{equation}
  \label{statc}
  C_i^s = \frac{G_i \tau_i(\chi)}{a}\,.
\end{equation}
This yields (using again $f_\eta^2 e^\phi=1$)
\begin{equation}
\label{statg} \frac{C_1^s}{C_2^s} = \frac{G_1}{G_2}
\frac{\tau_1(\chi)}{\tau_2(\chi)} = \frac{G_1}{G_2}
\exp(\theta(1-2\chi)) \equiv N_G(\chi)
\end{equation}
in the stationary limit. The function $N_G(\chi)$ is also shown in
Fig.~\ref{F:LSA:1}. Both the functions $N_\chi(\chi)$ and $N_G(\chi)$ intersect
at $\chi = \chi^\text{st}$ given by
\begin{equation}\label{sec:LSA:1}
    \chi^\text{st} = \frac{G_1}{G_1+ G_2} \equiv \frac{G_1}{G}
\end{equation}
thereby defining the total incoming flux $G = G_1 + G_2$. It defines the
stationary point of the present system for which both Eq.~\eqref{Diffusion} and
Eq.~\eqref{chidot} have vanishing time-derivatives.

Now we discuss some immediate consequences of the properties of the
functions $N_\chi(\chi)$ and $N_G(\chi)$. For this purpose we
consider the $(\chi,C_1^s/C_2^s)$-plane. { These functions separate
different regions of the phase plane. Naturally, a point on this
plane does not describe the complete system configuration
 because the diffusion is reduced to $C(z=0)$.} For
$C_1^s/C_2^s > N_\chi(\chi)$ the {time} derivative of $\chi(t)$ is
positive and vice versa. This means that a system above
$N_\chi(\chi)$ is driven toward larger $\chi$, whereas a system
below behaves oppositely.

In case of a standard relaxation oscillator the equation for
$\dot{\chi}$ would be complemented by an equation for $(d/d t)
(C_1^s/C_2^s)$. Such an equation does not exist for the present
model. {  However, an implicit time evolution of $(C_1^s/C_2^s)$ is
expressed by Eq. \eqref{statg}. When $(d/dt) \chi(t)$ is very small
the diffusion field can adjust to the boundary conditions. This way
the ratio $C_1^s/C_2^s$ will be adjusted until  Eq. \eqref{statg} is
fulfilled. In particular, $C_1^s/C_2^s$ will decrease for
$(C_1^s/C_2^s) > N_G(\chi)$ and vice versa. This is reflected by the
arrows in Fig.\ref{F:LSA:1}.}

{ The simplest form of a possible time evolution giving rise to
oscillatory behavior is shown in Fig. \ref{F:LSA:1}.} If there is a
time-scale separation between the $\chi$-variations and the
variations of the surface concentrations $C_i^s$ the system would
move from the left maximum of $N_\chi(\chi)$ to the right { until
$\dot{\chi} \approx (d/d t) (C_1^s/C_2^s)$}. Then the system has
time to adapt $(C_1^s/C_2^s)$ to the present value of $\chi$ thereby
moving down along the $N_\chi(\chi)$-curve. { Once the the minimum
is reached $(d/d t) (C_1^s/C_2^s)$ will drive the trajectory away
from $N_\chi(\chi)$ quickly resulting in $|\dot{\chi}| \gg |(d/d t)
(C_1^s/C_2^s)|$ which will move it towards the left branch of
$N_\chi(\chi)$. Then the second half of the oscillation may start.
As will be shown below via numerical simulations the actual behavior
is somewhat different. In any event, the general possibility of
oscillatory behavior requires a non-monotonous behavior of
$N_\chi(\chi)$, i.e. $\theta
> \theta_c= 2$. However, the linear stability analysis will reveal that
this condition is not sufficient.}

\subsection{Linear stability analysis}
The stability of the stationary crystal growth is analyzed with respect to
infinitesimal perturbations of the solute distribution in the solution and the
solid state composition around $\chi = \chi^\text{st}$ and the corresponding
values of $C_i^s$, given by Eq.~\eqref{statc} which will be denoted
$C_i^\text{st}$. We choose
\begin{subequations}\label{sec:LSA:4}
\begin{align}
     \label{sec:LSA:4a}
    \delta C_i(z,t) & = \delta C_i^s \exp\left(\gamma t - p_i z\right)
 \\
 \intertext{and}
     \label{sec:LSA:4b}
    \delta \chi(t) & = \delta \chi \exp\left(\gamma t\right).
\end{align}
\end{subequations}
Here $\delta C_i^s$ and $\delta \chi$ are the amplitudes whereas the complex
wave number $p_i = \operatorname{Re}p_i + \, \mathrm{i} \operatorname{Im}p_i $
describes  the decay of the concentration perturbations above the crystal
surface. It requires $\operatorname{Re}p_i > 0$. The instability arises when
the real part of the perturbation increment $\gamma$ becomes positive. The
chosen form~\eqref{sec:LSA:4} is compatible with the time-evolution close to
the stationary point.

The infinitesimal form of Eq.~\eqref{chidot} is of primary
interest. A short calculation gives
\begin{equation}
\label{gamma1} \frac{\gamma}{a^2G} = -1+ \chi^\text{st}(1-\chi^\text{st})\left
[2\theta + \frac{\delta C_1}{C_1^\text{st} \delta \chi} - \frac{\delta
C_2}{C_2^\text{st} \delta \chi} \right ].
\end{equation}
The relevant ingredient $\delta C_i^s/(C_i^\text{st}\delta \chi)$ can be
obtained from the boundary condition~\eqref{boundaryfin} as
\begin{equation}
\label{delta2} \frac{\delta C_i^s}{C_i^\text{st} \delta \chi} = \frac{a\partial
\ln \tau_i/\partial \chi}{a + D_i p_i \tau_i(\chi^\text{st})}.
\end{equation}
Finally, from Eq.~\eqref{Diffusion} one has
\begin{equation}\label{EVQ:8}
    \gamma = D_1p_1^2 = D_2p_2^2\,,
\end{equation}
i.e. the relation between the $p_i$ and $\gamma$.
Combining Eqs.~\eqref{gamma1}, \eqref{delta2} and \eqref{EVQ:8} and using the
specific dependencies~\eqref{taudef} we finally get
\begin{multline}\label{EVQ:11}
    \frac{\gamma}{a^2G} = -1 + \chi^\text{st}(1-\chi^\text{st})
\\
    \times\bigg[
        (\theta + \phi)
     \frac{D_1\tau_1 p_1}{D_1\tau_1 p_1 + a}
     +  (\theta - \phi)
    \frac{D_2\tau_2 p_2}{D_2\tau_2 p_2 + a}
        \bigg]\,.
\end{multline}

How does the nature of the stability depend on the total flux $G$?
One always has  $C_i^\text{st} \propto G$. Furthermore, in the limit
of large $G$ one also obtains $\gamma \propto G$ and, using
Eq.~\eqref{EVQ:8}, $p_i \propto \sqrt{G}$. As a consequence one has
$\delta C_i \propto C_i / p_i \propto \sqrt{G}$. Thus,
Eq.~\eqref{gamma1} boils down to
\begin{equation}
\label{gamma2} \gamma = {a^2G} [ -1+ \chi^\text{st}(1-\chi^\text{st})
2\theta]\,.
\end{equation}
One has $\gamma > 0$ exactly when $\chi^\text{st} \in (\chi_-,\chi_+)$, i.e.
$\chi^\text{st}$ is on the unstable branch of $N_\chi(\chi)$. In contrast, for
$G \rightarrow 0$ one has $\gamma = -a^2G < 0$. This can be seen from
Eq.~\eqref{EVQ:11} because in this limit also $p_i \rightarrow 0$. Thus there
exists a critical flux $G_c$ such that $\operatorname{Re}\gamma = 0$ for
$G=G_c$  and $\operatorname{Re}\gamma > 0$ for $G
> G_c$. Since $\gamma = p_i = 0$ cannot be a solution from
Eq.~\eqref{EVQ:11} the disappearance of $\operatorname{Re}\gamma$ implies that
$\gamma$ is purely imaginary.

Our goal is, first, to determine $G_c$ explicitly and, second, to
understand its dependence on the model parameters on a more
qualitative level, i.e. $G_c = G_c(\chi\mid\theta,\phi,\ldots)$.

\subsection{Exact solution}
From Eq.~\eqref{EVQ:11} the critical value $G_c$ can be determined. As shown in
the Appendix the critical value $G_c$ in parametric form is given by the
expression
\begin{multline}\label{sec:LSA:5}
    G_c = \frac1{\sqrt{D_1D_2}\tau^2}
    \frac{e^{-\phi(1-2\chi)-\theta}\zeta}{\sqrt2\chi(1-\chi)}
\\
   \times \bigg[
     \frac{(\theta+\phi)\Delta}
              {(\sqrt2\zeta \Delta + 1)^2 +1}
    +
     \frac{(\theta-\phi)/\Delta}
              {(\sqrt2\zeta/ \Delta + 1)^2 + 1}
    \bigg]^{-1}\,.
\end{multline}
Here the variable $\zeta$ is the root of the equation
\begin{equation}\label{sec:LSA:6}
       (\theta + \phi) \Psi\big(\sqrt2 \zeta\Delta\big)
  +    (\theta - \phi) \Psi\big(\sqrt2 \zeta/\Delta\big)
     = \frac1{\chi(1-\chi)}\,,
\end{equation}
where the function $\Psi(x)$ is defined as
\begin{equation}\label{sec:LSA:7}
    \Psi(x) = \frac{x(x+1)}{(x+1)^2+1}
\end{equation}
and the function $\Delta(\chi)$ is
\begin{align}
 \label{sec:LSA:8}
    \Delta(\chi)  & = \Delta_{0.5}e^{\theta(1-2\chi)/2}
\\
\intertext{with the prefactor $\Delta_{0.5}$ given by the expression}
 \label{sec:LSA:9}
     \Delta_{0.5}^2  & = \sqrt{\frac{D_1}{D_2}}f_\eta^{-2}
    e^{-\phi}\,.
\end{align}

Since $\phi \leq \theta$,  both terms on the left-hand side of
Eq.~\eqref{sec:LSA:6} are non-negative increasing functions of the
argument $\zeta$. The maximum of their sum is equal to $2\theta$
whereas the minimum of the right-hand side at $\chi = 0.5$ is equal
to 4. So, we obtain again the condition $\theta > \theta_c$
necessary for instable behavior. In this case for
$2\theta\chi(1-\chi)
> 1$ equation~\eqref{sec:LSA:6} possesses only one root which
together with \eqref{sec:LSA:5} determine the lower boundary $G_c$
of the instability region. In the close vicinity of the threshold,
$0 <2\theta\chi(1-\chi) - 1\ll 1$ (for $\chi\approx 0.5$) this
equation can be solved analytically, giving the expression
\begin{gather}
 \label{sec:LSA:10}
    \zeta  \approx\frac{\left[
        (\theta + \phi)/{\Delta_{0.5}} + (\theta - \phi)\Delta_{0.5}
    \right]}
    {4\sqrt{2}[2\theta\chi(1-\chi)-1]}
\\
\intertext{and thus}
 \label{sec:LSA:11}
    G_c \approx \frac{e^{-2}}{32\sqrt{D_1D_2}\tau^2}\,
    \frac{[
        (\theta + \phi)/{\Delta_{0.5}} + (\theta - \phi)\Delta_{0.5}
    ]^2}
    {[2\theta\chi(1-\chi)-1]^3}\,.
\end{gather}

For the system with $\phi = 0$ the asymmetry of the solubilities substantially
increases  the critical diffusion flux $G_c$. Indeed, since the species
diffusivities in solutions are typically of the same order, $D_1\sim D_2$, the
species solubility difference reflected in the coefficient $f_\eta\ll 1$ or
$f_\eta\gg 1$ matches (see Eq.\eqref{sec:LSA:9}) the inequality $\Delta_{0.5}
\gg 1$ or $\Delta_{0.5} \ll 1$, respectively. For $\phi\approx\theta$ the term
$(\theta-\phi)\Delta_{0.5}$ in \eqref{sec:LSA:11} can be neglected and an
increase of $\Delta_{0.5}$ gives rise to a remarkable decrease in $G_c$.

\begin{figure}
\begin{center}
\includegraphics[width=\columnwidth]{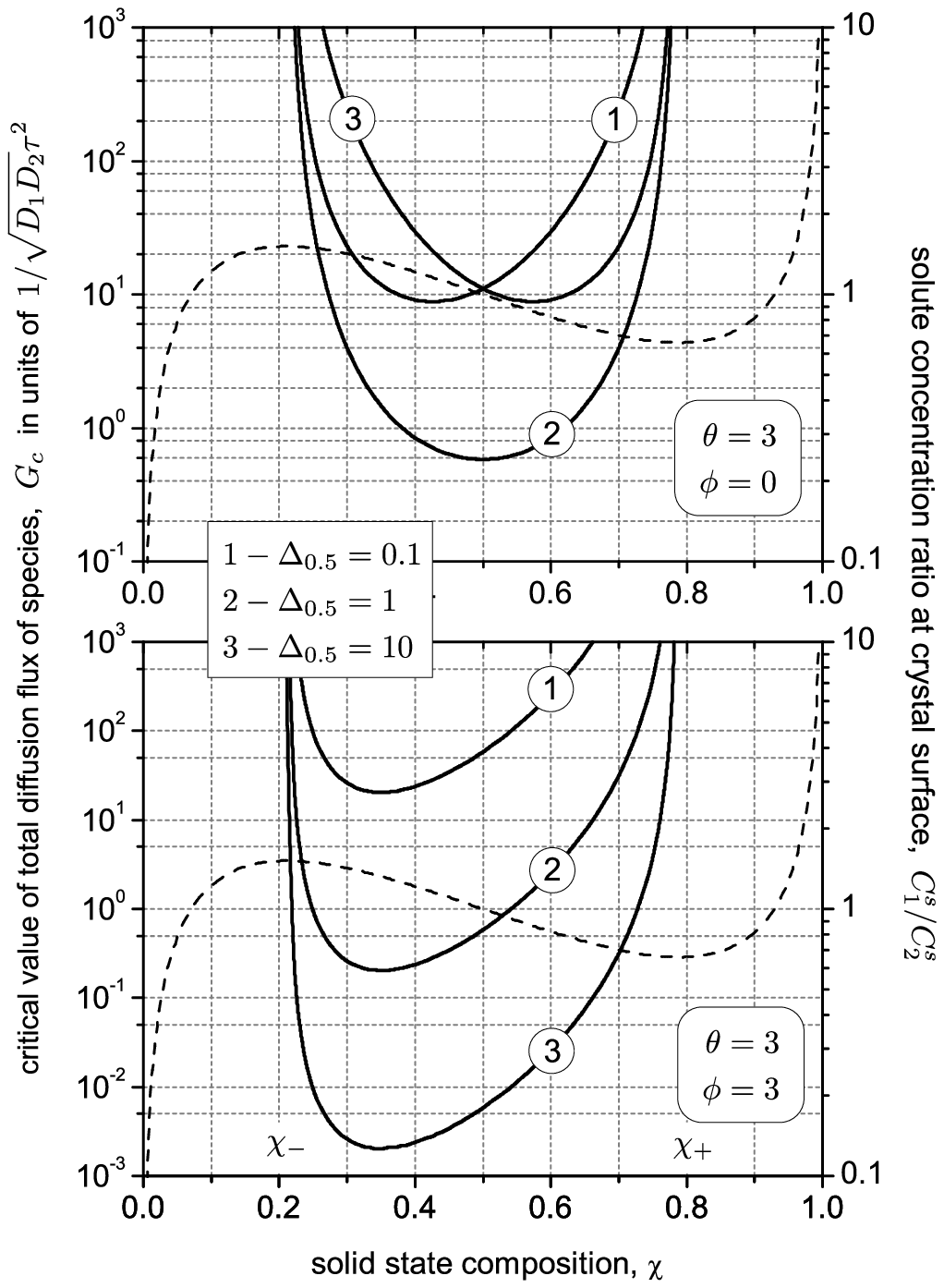}
\end{center}
\caption{The critical value of $G_c$ vs. $\chi$ for $\theta=3$ and
some different
 $\Delta_{0.5}$ representing the difference in the
species solubilities. The upper frame corresponds to a symmetric
system with  $\phi= 0$, the lower one to a strongly asymmetrical one
with $\phi = 3$. The dashed curves correspond to the function
$N_\chi(\chi)$.
\label{F:LSA:2}}
\end{figure}

Equation~\eqref{sec:LSA:6} was solved numerically to analyze the
system behavior far from the threshold $\theta_c = 2$. For $\theta =
3$ the results for the $\chi$-dependence of $G_c$ are presented in
Fig.~\ref{F:LSA:2}. In addition we have included $N_\chi(\chi)$ in
Fig.~\ref{F:LSA:2}. The upper frame exhibits the results for $\phi =
0$, whereas the lower one contains the strongly asymmetrical case
with $\phi = \theta = 3$.

The symmetrical system exhibits minimal $G_c$ when the species have
the same solubility. A difference in solubility as reflected by the
change of $\Delta_{0.5}$ by a factor of ten causes $G_c$ to increase
by a similar factor. For such  $\Delta_{0.5}$ the $G_c(\chi)$-curves
become asymmetrical with respect to $\chi = 0.5$. Exactly this case
should be characterized by the instability forming the limit cycle
constructed in Fig.~\ref{F:LSA:1} following the standard notions of
relaxation oscillations.

The system behavior for $\phi\sim \theta$ is distinctly different as
can be seen in the lower fragment of Fig.~\ref{F:LSA:2}. In
particular, for $\phi = 3$ a ten-fold increase in $\Delta_{0.5}$
induces a hundred-fold drop in $G_c$. It should be noted that
$\Delta_{0.5}=1$ corresponds to $\eta =\phi/2$ which is non-zero
here. Furthermore, for $\Delta_{0.5} =10$ the dependence $G_c(\chi)$
is highly asymmetrical and passes many orders of magnitude. For
large values of $\Delta_{0.5}$, the left part of the decreasing
branch of $N_\chi(\chi)$ can be unstable whereas the right half can
be stable. In this case the limit cycle of the developed
oscillations deviates substantially from the classical form.

\subsection{Instability mechanism: qualitative description}

It is possible to obtain a better understanding of how the degree of
instability depends on the system parameters. We use sufficiently
large $G$ such that $a \ll |D_i p_i \tau_i|$. The values of $\tau_i$
are always analyzed at $\chi = \chi^\text{st}$. For reasons of
simplicity we also assume $D_1 = D_2 = D$ (implying $p \equiv p_1 =
p_2$). Then we can rewrite Eq.~\eqref{delta2} as
\begin{gather}
\frac{\delta C^s_i}{C_i^\text{st}  \delta \chi} = \frac{a}{p D}\frac{\partial
\ln\tau_i/\partial \chi}{\tau_i}\\
\intertext{thereby} \label{move} \frac{\delta C^s_1}{C_1^\text{st}  \delta
\chi} - \frac{\delta C^s_2}{C_2^\text{st}  \delta \chi} =
 \frac{a}{p{D}}\left[ -\frac{\theta}{\tau_1} -
\frac{\theta}{\tau_2}- \frac{\phi}{\tau_1}+ \frac{\phi}{\tau_2}
\right ].
\end{gather}
The cumulative effect of these terms gives rise to a decrease of the real part
of $\gamma$. Thus, concentration fluctuations always tend to stabilize the
stationary point. It is interesting to analyze the impact of the asymmetry
$\phi$. In what follows we will restrict ourselves to the case $\phi = \theta$,
where $\delta C_2 = 0$ and the damping is due to the concentration fluctuations
of species 1. Changing from $\phi = 0$ to $\phi = \theta$ the relevant term in
Eq.~\eqref{gamma1}, characterizing the concentration fluctuations and thus the
reduction of the real part of $\gamma$ changes from $-(1/\tau_1 + 1/\tau_2)$ to
$-2/\tau_1$ which corresponds to a change by a factor of $2/(1+\tau_1/\tau_2)$.
Note that
\begin{equation}
\tau_1/\tau_2 =
    \Delta_{0.5}^2\exp[\theta(1-2\chi^\text{st})]\,.
\end{equation}
When $\Delta_{0.5}\gg 1$ the result of the concentration
fluctuations is thus reduced which has a positive cumulative effect
on the instability onset and, as a result,  $G_c$ should be
decreased. Moreover, this ratio increases as $\chi$ decreases. This
rationalizes the asymmetry of the $G_c(\chi)$-dependence with
smaller values located on the left-hand side. Exactly these features
emerge from the solution of the exact equation~\eqref{EVQ:11}.

\section{Numerical analysis of the nonlinear dynamics}\label{NL}

When the growth becomes unstable nonstationary patterns in the solution and the
induced  pattern in the crystal bulk develop. In order to analyze their
characteristic properties the system of equations in Sect.~\ref{modelall} was
studied numerically. First, we introduced the spatial and temporal scales
$l_\text{sc}$ and $\tau_\text{sc}$
\begin{gather}
\label{NL:1}
    l_\text{sc} = \left(\frac{\sqrt{D_1D_2}\tau}{a^2}\right)a\,,\quad
    \tau_\text{sc} = \left(\frac{\sqrt{D_1D_2}\tau}{a^2}\right)\tau\,,
\\
\intertext{and, in addition, a parameter having the dimension of
concentration} \label{NL:2}
     C_\text{sc} = \frac1{a\sqrt{D_1D_2}\tau}\,.
\end{gather}
Then we rescaled time $t$ and spatial coordinate $z$ as well as the species
concentrations $C_i$ in these units
\begin{equation*}
    t\rightarrow \tau_\text{sc}\cdot t\,,\quad
    z\rightarrow l_\text{sc}\cdot z\,,\quad
    C_i \rightarrow C_\text{sc}\cdot C_i\,.
\end{equation*}
For the sake of simplicity we have kept the same designations for these
variables. The fluxes $G_i$, $G$ were also measured in units of
\begin{equation*}
    G_\text{sc} = \frac1{\sqrt{D_1D_2}\tau^2}\,,
\end{equation*}
namely,
\begin{equation}\label{NL:9}
    (G, G_i) \rightarrow G_\text{sc}\cdot (G, G_i)\,.
\end{equation}
Then the obtained system of dimensionless equations was integrated using the
Crank-Nicholson method. The temporal and spatial steps of integration as well
as the system size $L$ were chosen such that the dynamics be practically
independent of their particular values.

The following four specific cases were analyzed to demonstrate characteristic
features of the physical system. The first two ones are the symmetrical model
with parameters $\theta = 3$, $\phi = 0$, $\Delta_{0.5} = 1$  and
$\chi_\text{st} = 0.5$ at the initial stage. They differ in the total diffusion
flux $G = 1$ and $G=50$. The case with $G=1$ describes the system dynamics not
too far from the instability boundary $G_c \approx 0.6$  (see
Fig.~\ref{F:LSA:2}). Under these conditions the instability was expected to
demonstrate quasiharmonic behavior. The case $G = 50$ corresponds to the
substantially nonlinear stage of the instability.

The other pair of cases are the asymmetrical model with $\theta = 3$ and $\phi
= 3$ being actually a limit situation that can be considered accurately within
the present analysis. To single out the effects caused by nonzero values of
$\phi$ we restrict ourselves to $\Delta_\text{0.5} = 1$ and $\chi_\text{st} =
0.3$. Correspondingly, $G_c(\chi)$ is minimal for this value of $\chi$. The
values of $G$ are set to $0.5$ and $5$.

\begin{figure*}
\begin{center}
\includegraphics[width = 0.9\textwidth]{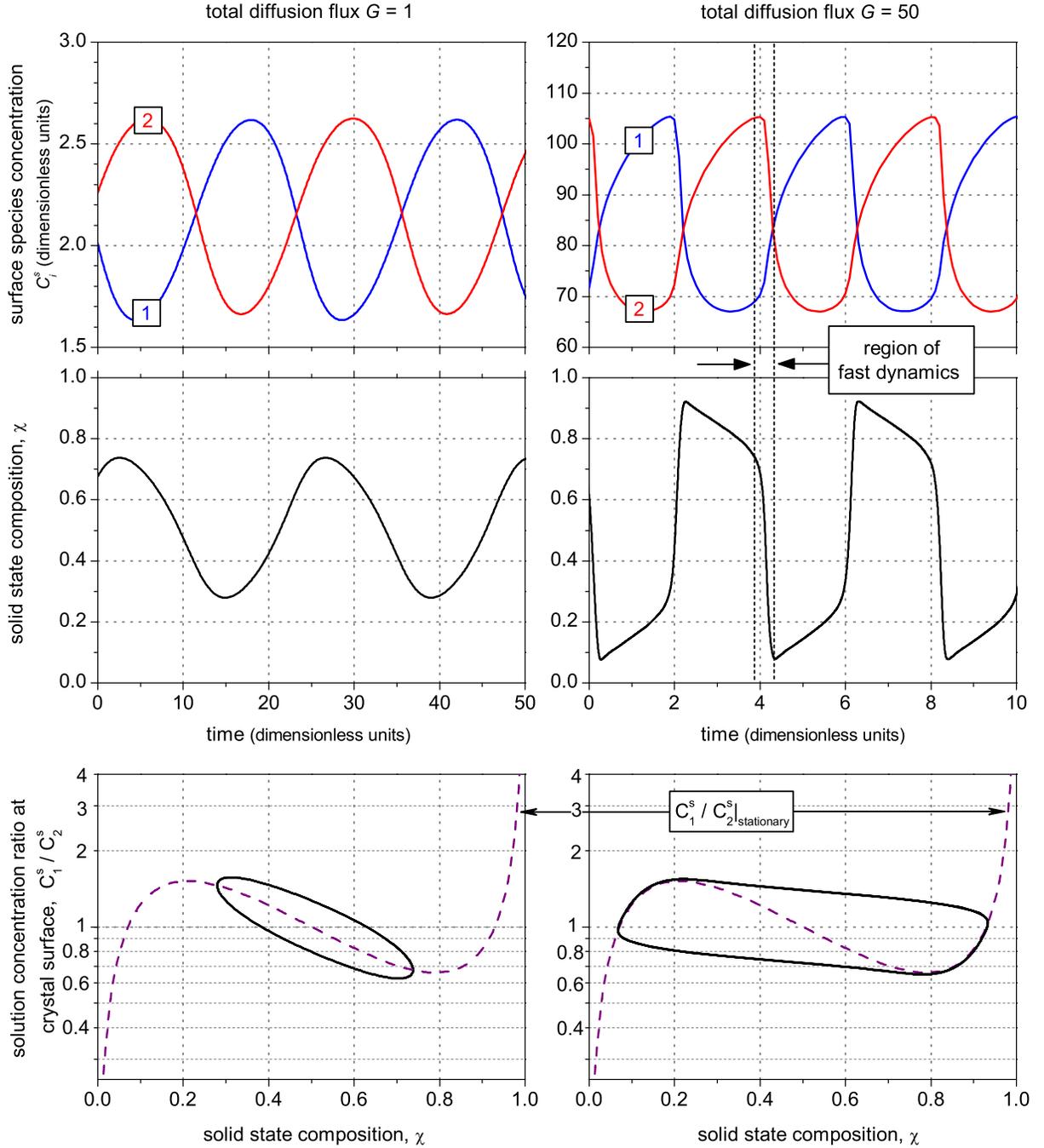}
\end{center}
\caption{Dynamics of the symmetrical system $\theta = 3$, $\phi =
0$, $\Delta_{0.5} = 1$, $\chi_\text{st} = 0.5$. The left column has
the results obtained for  $G = 1$ which is larger than $G_c = 0.6$ ,
see Fig.~\ref{F:LSA:2}. The right column exhibits the results for
$G=50$.} \label{F100}
\end{figure*}

\begin{figure*}
\begin{center}
\includegraphics[width = 0.9\textwidth]{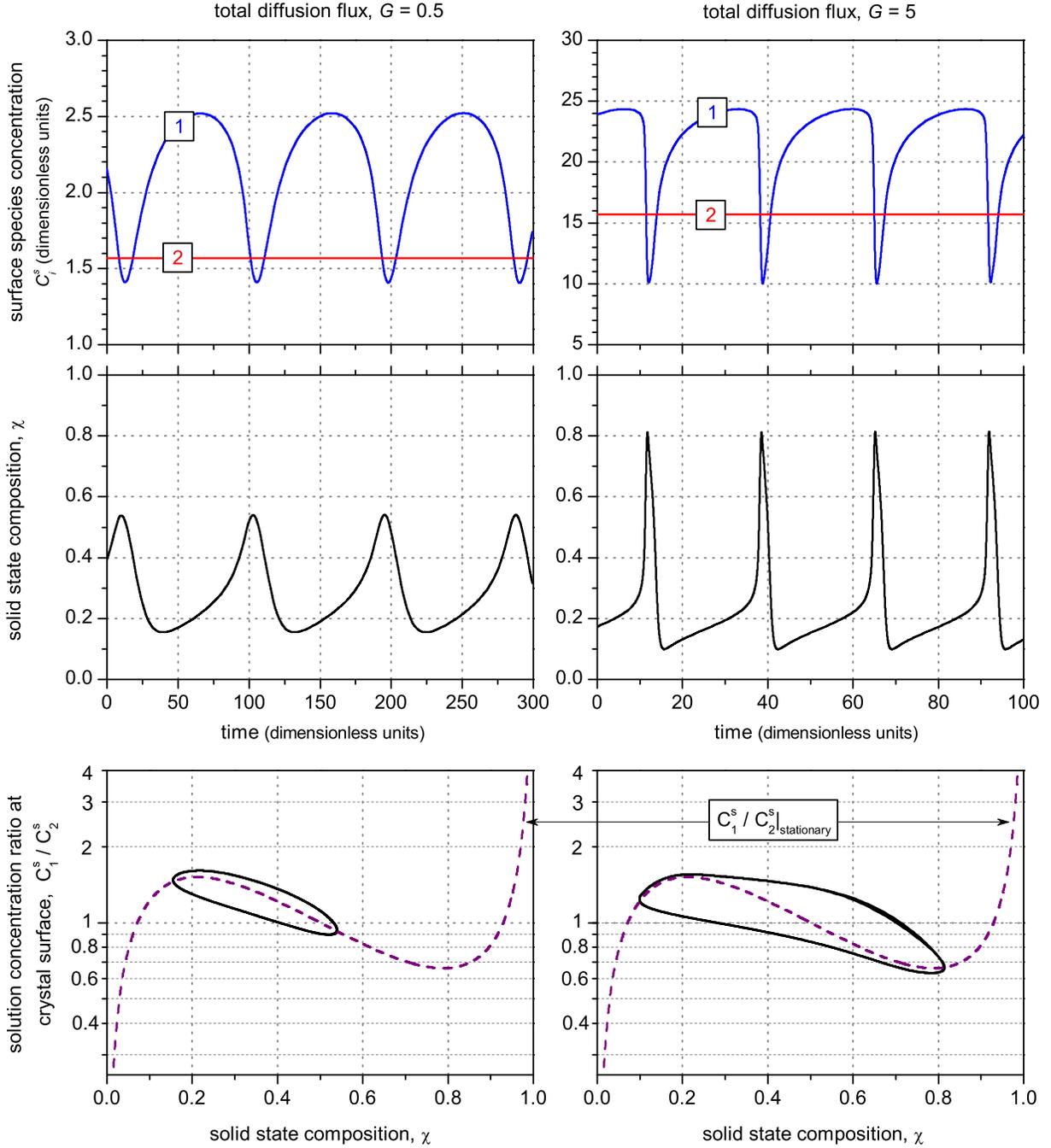}
\end{center}
\caption{Dynamics for the asymmetrical system $\theta = 3$, $\phi = 3$,
$\Delta_{0.5} = 1$, $\chi_\text{st} = 0.3$. The left column presents the data
obtained for the total diffusion flux $G = 0.5$, whereas the right one exhibits
the same data for $G=5$. } \label{F110}
\end{figure*}

\begin{figure*}
\begin{center}
\includegraphics[width = 0.9\textwidth]{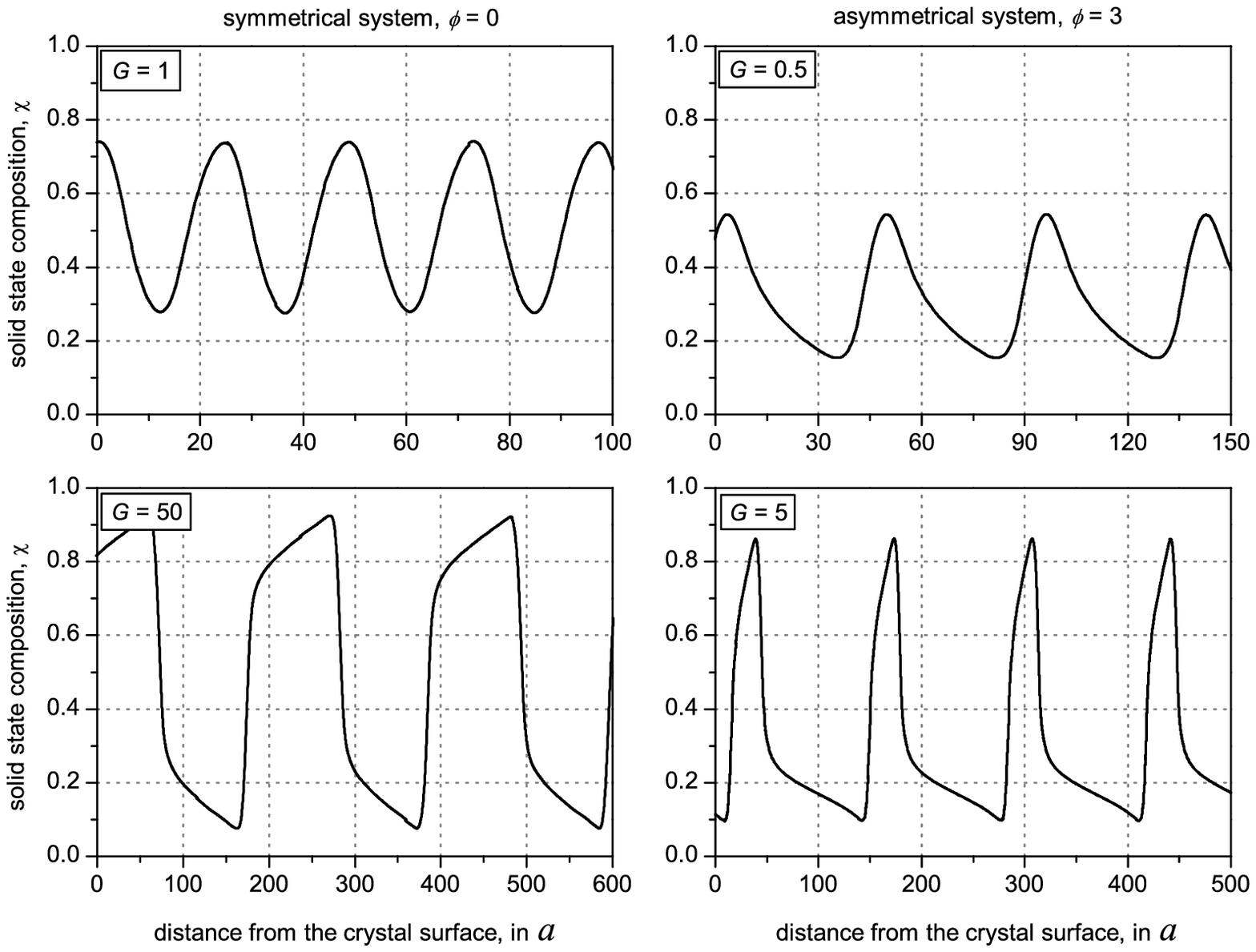}
\end{center}
\caption{Spatial patterns formed in the growing crystal.} \label{patterns}
\end{figure*}

Figure~\ref{F100} visualizes the surface dynamics for the first two cases. As
seen in the frames of its left column, the oscillations of $C_i^s$ and $\chi$
are rather harmonic as  expected. The resulting limit cycle is shown in the
lower left frame. Again, as expected, this cycle is of elliptical form located
along the unstable branch of the nullcline $N_\chi(\chi)$. Even for $G = 1$
which is close to $G_c \approx 0.6$ the oscillations are already determined by
the nullcline.

For $G = 50$ the dynamics become relaxation-like and $\chi$ plays the role of
the fast variable. As seen in the right column of Fig.~\ref{F100}, the time
pattern $\chi(t)$ consists of a sequence of slow motion fragments joined by
rather sharp jumps. In agreement with  classical relaxation oscillations the
fragments of slow dynamics correspond to the system motion along  stable
branches of $N_\chi(\chi)$ whereas the sharp jumps describe the fast
transitions between these branches. However, the surface species concentrations
exhibit anomalous behavior from the standpoint of classic relaxation
oscillations. Most importantly, $C_1^s/C_2^s$ display some remarkable
variations during the fast dynamics phase of $\chi$. This is still reminiscent
of the result of the linear stability analysis where $\delta C_i^s \propto
\delta \chi$. As a result the lines of the limit cycle connecting the
increasing branches of the nullcline $N_\chi(\chi)$ are not horizontal but
inclined to the $\chi$-axis at a certain visible angle. It should be noted that
this inclination is not a standard consequence of the finite ratio between the
time scales of fast and slow dynamics because, otherwise, the form of these
curves would deviate form the straight lines substantially. It is due to the
existence of the simultaneous change in both the species concentrations during
the period of fast motion.

The asymmetrical system with $\theta = 3$, $\phi = 3$ is depicted in
Fig.~\ref{F110} for $G = 0.5$ and $G = 5$. The basic feature distinguishing
this system from the symmetrical one is the strong asymmetry of the critical
diffusion flux with respect to $\chi = 0.5$. In particular, as shown in
Fig.~\ref{F:LSA:2} for $G = 0.5$ only the left half of the decreasing branch of
the nullcline $N_\chi(\chi)$ is unstable whereas for $G = 5$ the instability
region is located approximately within the boundaries $0.2<\chi<0.6$. The
resulting limit cycles are shown in the lower frames of Fig.~\ref{F110}. For $G
= 0.5$ the stationary point is not too far from the instability boundary and
the limit cycle is located in a rather narrow neighborhood of the unstable
branch of $N_\chi(\chi)$. For $G = 5$ the separation of the system dynamics
into the slow and fast motion fragments becomes pronounced. In this case the
limit cycle embraces even an increasing branch of the $N_\chi(\chi)$. However,
its lower part after passing the minimum near $\chi = 0.8$ continues to follow
the formerly unstable fragment of the decreasing nullcline branch until it
reaches the instability boundary at $\chi\approx 0.6$. Only after passing this
point it leaves the branch and jumps to the opposite stable branch of
$N_\chi(\chi)$. This effect of ``adhesion'' to the unstable branch of
$N_\chi(\chi)$ is of another nature than the well known ``French duck''
fragment of the limit cycle for standard relaxation oscillations (see, e.g.,
Ref.~\cite{Izhikevich}), caused by the close proximity of the stationary point
to the extreme of the  nullcline.

The two time patterns of the observed oscillations deviate substantially in
shape both from the quasiharmonic oscillations and relaxation oscillations.
Especially for the flux $G = 5$ the found pattern looks like a sequence of
pronounced spikes joined by fragments of slow motion along the left decreasing
branch of $N_\chi(\chi)$.

The resulting crystal profiles are shown in Fig.~\ref{patterns}. Basically, the
profiles are scaled mirror images of the time-dependent $\chi$ as shown in
Figs.~\ref{F100} and \ref{F110}, because the growth velocity only weakly varies
with time. As should be expected the spatial oscillation period is orders or
magnitude larger than the atomic scale.

\section{Discussion}
We have presented a 1D model for OZ. The growth rate as the central non-linear
coupling term between the solution and the crystal was derived based on
layer-by-layer crystal growth mechanisms. For this purpose particle adsorption,
surface diffusion and finally desorption or inclusion processes at the steps
are taken into account. This way, adsorbed particles do not only experience the
local environment of their adsorption site, but a much more averaged one
depending strongly on the composition of the crystal surface. This is an
essential feature of the present model, because together with volumetric
species diffusion it may provide the synchronization effect necessary to
successfully describe the experimental findings in more than one dimension.

Our model (Eqs.\eqref{Diffusion}-\eqref{chidot}) differs in several respects
from the existing model of L'Heureux \textit{et al.}
\cite{PhanMathMod,KatsevNoise}. (i) The most essential difference is the choice
of the growth mechanism, i.e. the definition of the growth rate $r_i(\chi)$. In
Refs.~\cite{PhanMathMod,KatsevNoise} a phenomenological equation based on local
atom adsorption \cite{MarkovBuch} is formulated. It is one of the simplest
polynomial expression for $r_i(\chi)$ which displays nonlinear behavior. (ii)
The boundary condition at $z=0$ used in \cite{PhanMathMod,KatsevNoise} for
crystal growth from solution can be written as
\begin{equation}
\label{boundaryheureux} D_i\partial C_i/\partial z = \chi_i (r_1 +
r_2)
\end{equation}
in our notation. Here one uses $\chi_1 = 1 - \chi_2 = \chi$. It is
identical to Eq.~\eqref{boundaryfin} if (d/d$t) \chi(t) = 0$. But as
soon as $\chi$ changes with time both boundary conditions are not
identical. Actually, to derive Eq.~\eqref{boundaryheureux} one can
formulate the mass balance in a newly generated crystal layer. Then
Eq.~\eqref{boundaryheureux} only emerges if the $\chi$-value in the
layer at some time, determining the nature of the growth rates
$r_i(\chi)$ during the growth of the new layer is identical to the
$\chi$-value in the resulting new layer. In general this is not the
case. Thus we feel that Eq.~\eqref{boundaryfin} is more generally
valid although from a practical point of view no essential
difference should be present. (iii) We specify the gradient at $z=L$
whereas in \cite{PhanMathMod,KatsevNoise} the concentration is
given. Our choice is conceptually more simple because the value of
$L$ does not enter the calculations (see e.g. Eq.~\eqref{EVQ:6}).
However, the underlying physical picture does not change whether the
gradient or the concentration itself are fixed as long as $L$ is
sufficiently large. (iv) Another distinctive difference is the
explicit calculation of the diffusion field in the solution above
the crystal. We do not apply the boundary layer approximation,
regarding a crystal growing through a supersaturated solution
\cite{PhanMathMod}, but we neglect the spatial growth with respect
to the diffusion. This way, we obtain a source-sink system with a
diffusion field that has to be treated explicitly, because the
interaction of particle accumulation or depletion on the solution
side and the autocatalytic growth on the crystal side is essential
for the existence of OZ. In addition, we perform the stability
analysis without further approximations. This may be essential
because the oscillatory dynamics can stem from the counteraction of
strong ``forces'' mutually compensating each other at the first
approximation. { We would like to mention that in \cite{KatsevNoise}
the system of equations has been numerically solved without invoking
the boundary layer approximation. On a qualitative level similar
results were obtained as compared to the analytical treatment within
the boundary layer approximation}. { (v) The surface roughness
parameter, used in \cite{PhanMathMod}, is not necessary in the
present analysis.}

The proposed growth mechanism is valid for $a\ll l_s\ll l$. In the
limit of $l_s\approx a$ incorporation would be governed by the
local crystal composition, again. In case of $l_s\approx l$
desorption can be neglected and nearly every adatom will be
incorporated regardless of its type and $\chi$. This would result
in crystals exhibiting the stoichiometric compositions
\cite{PrietoZoningComp} of the influx.

{ The present model cannot exhibit the  bistability found in
\cite{PhanMathMod}, because the composition of the only stationary
point is determined by the influx ratio. Any crystal growth with a
composition different from this ratio will result in a buildup of
the currently ``disfavored'' component until its growth rate
increases. Thus the oscillations have to revolve around or run into
this fixed point. This corresponds well to the model picture, where
a bistability is only possible if either the supersaturation is not
high enough with respect to the minor component or if complementary
crystals grow in close vicinity.}

The time scales of diffusion and crystal growth define the qualitative
behavior of the oscillations. At very low concentrations the
crystal growth rate and consequently the changes in crystal
composition are very slow. The diffusive processes on the other
hand are fast enough to counteract this and no oscillatory
behavior is observed. With rising solute concentrations the growth
rate increases as well, whereas the characteristic speed of
diffusion stays constant resulting in soft transitions and
sinusoidal oscillations. At even higher concentrations, the speed
of crystal composition change supercedes the diffuse reaction of
the solution by many orders, creating sharp transitions, closely
following $N_\chi(\chi)$. This qualitative description  coincides
with the results from the linear stability analysis At low influx
it is stable possessing two imaginary eigenvalues with negative
real parts. At $G_c$ the real part changes sign and the system
becomes unstable.

Finally, the numerical  calculation of asymmetrical systems show
the stabilization of a formerly unstable branch of $N_\chi(\chi)$.
If the fixed point is shifted far enough onto the stable branch,
the trajectories slowly approach it along the nullcline, but upon
nearly reaching it are directed away with the onset of new
oscillations. The amplitude of these oscillations can have
continuously growing character or they can exhibit certain
characteristic values. This might be the onset of frequency
doubling and provide a route into chaotic behavior of the system.
This however, is beyond the scope of the present manuscript.

\section{Conclusion}
The presented boundary  reaction diffusion model, describing OZ of
crystals grown from solution. {Thereby it applies a mechanism
including surface diffusion, which can be derived from microscopic
properties.  OZ arises due to diffusive build up of the disfavored
component above the crystal until the autocatalytic growth process
in combination with the large population turn the crystal
composition in favor of the formerly disfavored species. A linear
stability analysis was carried out without further approximations.}
The results correspond to the experimental findings of a
supersaturation threshold before the onset of OZ. Numerical
simulations of asymmetric cases show a stabilization of previously
unstable parts of the system.

\begin{acknowledgments}
Felix Kalischewski would like to thank the "Fonds der Chemischen Industrie" and
the "NRW Graduate School of Chemistry" for financial support. Ihor Lubashevsky
appreciates the financial support of the SFB 458 and the University of
M\"unster as well as the partial support of RFBR Grant 04-02-81059. We thank
Prof. Putnis and Prof. Purwins for helpful discussions.
\end{acknowledgments}

\appendix

\section*{Eigenvalue equation for the boundary-reaction diffusion
model} \label{app:EVQ}

Let us convert the system of equations~\eqref{EVQ:8} and \eqref{EVQ:11} into a
form more appropriate for its analysis.
 By virtue of \eqref{EVQ:8} both the wave numbers $p_i$
have the same argument $\psi \in (-\pi/2;\pi/2)$, i.e. $p_i = |p_i|
\exp(i\psi)$ whose twice value gives the argument of $\gamma$ and their
absolute values obey the equality $|p_i| = \big(|\gamma|/D_i\big)^{1/2}$.

A new variable $\zeta >0$ and parameter $\Delta$ introduced as follows
\begin{align}
    \label{EVQ:12a}
    \zeta^2 & = \frac{\sqrt{D_1D_2}\tau_1\tau_2}{a^2}|\gamma|
    = \frac{\sqrt{D_1D_2}\tau^2}{a^2}
    e^{\phi(1-2\chi)+\theta}
    |\gamma|\,,
\\
    \label{EVQ:12b}
    \Delta ^2 & = \frac{\sqrt{D_1}\tau_1}{\sqrt{D_2}\tau_2}
    =\sqrt{\frac{D_1}{D_2}}f_\eta^{-2}
    e^{-\phi +\theta(1-2\chi)}
\end{align}
enable us, first, to write
\begin{equation}\label{EVQ:13}
    D_1\tau_1 |p_1| = a \zeta \Delta \,,\qquad D_2\tau_2 |p_2| = a\zeta/\Delta
\end{equation}
and, then, to represent equation~\eqref{EVQ:11} in the form
\begin{multline}\label{EVQ:14}
    \frac{\zeta^2}{g}\,e^{\mathrm{i}2\psi} = -1 + \chi(1-\chi)
\\
    \times\bigg[(\theta + \phi)
     \frac{(\zeta \Delta) e^{\mathrm{i}\psi}}{(\zeta \Delta) e^{\mathrm{i}\psi} + 1}
     +  (\theta - \phi)
     \frac{(\zeta/\Delta) e^{\mathrm{i}\psi}}{(\zeta /\Delta) e^{\mathrm{i}\psi} + 1}
        \bigg]\,,
\end{multline}
where the parameter $g$ stands for the dimensionless species flux casing the
crystal growth
\begin{equation}\label{EVQ:15}
    g = \sqrt{D_1D_2}\tau_1\tau_2\, G =
    \sqrt{D_1D_2}\tau^2
    e^{\phi(1-2\chi)+\theta}G\,.
\end{equation}
Finally the split of equality~\eqref{EVQ:14} into the real and imaginary parts
yields the desired coupled equations specifying actually the eigenvalue
$\gamma$
\begin{widetext}
\begin{subequations}\label{EVQ:16}
\begin{align}
  \label{EVQ:16r}
  \frac{\zeta^2}{g}\cos(2\psi) & = -1  + \chi(1-\chi)
    \bigg[ (\theta + \phi)
     \frac{(\zeta \Delta )(\zeta \Delta +\cos\psi)}
              {(\zeta \Delta + \cos\psi)^2 + \sin^2\psi}
    +
     (\theta - \phi)
     \frac{(\zeta /\Delta)(\zeta /\Delta+\cos\psi)}
              {(\zeta/ \Delta + \cos\psi)^2 + \sin^2\psi}
    \bigg]\,,
\\
\label{EVQ:16i}
    \frac{\zeta^2}{g}\sin(2\psi)& = \chi(1-\chi)\sin\psi
    \bigg[(\theta+\phi)
     \frac{(\zeta \Delta)}
              {(\zeta \Delta + \cos\psi)^2 + \sin^2\psi}
    +
     (\theta-\phi)
     \frac{(\zeta /\Delta)}
              {(\zeta/ \Delta + \cos\psi)^2 + \sin^2\psi}
    \bigg]\,.
\end{align}
\end{subequations}
\end{widetext}
As follows from equations~\eqref{EVQ:16} the value $\zeta = 0$ is not a root of
this system. So the considered instability is to arise through the real part of
$\gamma$ changing the sign with its imaginary part having some finite value.
Therefore to find the instability boundary $\operatorname{Re}\gamma = 0$ we can
set $\psi = \pi/4$ in the given equations. In this way the former
equation~\eqref{EVQ:16r} converts into one specifying the value of $\zeta$ for
chosen values of the parameters $\chi$, $\theta$, $\phi$, and $\Delta$
\begin{widetext}
\begin{gather}
  \label{EVQ:17}
    (\theta + \phi)
     \frac{(\sqrt2\zeta \Delta )(\sqrt2\zeta \Delta +1)}
              {(\sqrt2\zeta \Delta + 1)^2 + 1}
    +
     (\theta - \phi)
     \frac{(\sqrt2\zeta /\Delta)(\sqrt2\zeta /\Delta+1)}
              {(\sqrt2\zeta/ \Delta + 1)^2 + 1}
     = \frac1{\chi(1-\chi)}\,,
\\
\intertext{whereas the latter equation determines the critical value of the
diffusion flux $g_c$}
\label{EVQ:16ii}
    g_c = \frac{\zeta}{\sqrt2\chi(1-\chi)}
    \bigg[(\theta+\phi)
     \frac{\Delta}
              {(\sqrt2\zeta \Delta + 1)^2 +1}
    +
     (\theta-\phi)
     \frac{1/\Delta}
              {(\sqrt2\zeta/ \Delta + 1)^2 + 1}
    \bigg]^{-1}\,.
\end{gather}
Returning to the dimensional variables we get expressions~\eqref{sec:LSA:5} and
\eqref{sec:LSA:6}.
\end{widetext}


\end{document}